\documentclass[12pt, letterpaper]{article}

\pdfoutput=1

\usepackage{arxiv}
\usepackage[utf8]{inputenc} 
\usepackage[T1]{fontenc}    
\usepackage{amsmath,amsfonts,amssymb,bm}
\usepackage{mathtools}
\usepackage{mathptmx}
\usepackage{newtxtext}
\usepackage{newtxmath}
\usepackage{hyperref}
\usepackage{caption}
\usepackage{float}
\usepackage{xcolor}
\usepackage{multirow}
\usepackage{subcaption,booktabs}
\usepackage{rotating}
\usepackage{adjustbox}
\usepackage{graphicx}
\usepackage{array}
\usepackage{environ}
\usepackage[square,sort,comma,numbers]{natbib}
\DeclareMathOperator*{\argmax}{arg\,max}

\title{Changepoint Methods in Climatology}

\author{
	Robert B Lund \\
	Department of Statistics \\
	The University of California, Santa Cruz \\
	Santa Cruz, CA 95064 \\
	~\\
	and \\
	~\\
	Xueheng Shi \\
	Department of Statistics \\
	The University of Nebraska-Lincoln \\
	Lincoln, NE 68588 \\
	~\\
	(\emph{submitted to CHANCE Magazine})
}

\begin{document}

\maketitle

\section{Introduction}

Changepoints are discontinuity times (abrupt changes) in a time-ordered sequence of data.  In climate settings, changepoints often occur when measuring stations are relocated or gauges are changed.  Moving a climate station even 100 yards, for example, can shift temperatures by several degrees, especially if the solar exposure of the new location differs. In the United States, first-order climate stations average roughly six relocations or gauge changes per century. In many instances, these change times are not known to users analyzing the record.  Changepoint times can also be triggered by natural causes when weather patterns shift.

A definition of a single changepoint for a random sequence $\{ X_t \}_{t=1}^N$ of length $N$ is an unknown time $\tau \in \{ 2, 3, \ldots, N \}$ such that the distribution of $X_t$ for $t \leq \tau$ is $F_{\rm Before}(\cdot)$, and shifts to $F_{\rm After}(\cdot)$ after the changepoint time when $t > \tau$.  There are many ways in which the marginal distribution can shift. Perhaps the simplest way allows the means of the series to shift at each changepoint time.  This mean shift case will become our focus later. Researchers have also studied shifts in process variabilities (volatilities) in finance and shifts in process autocovariances in speech recognition.

In the mean shift case, a regression model for the series is
\[
X_t= \mu_t +\epsilon_t,
\]
where $\mu_t=E[X_t]$ is the time $t$ mean of the series and $\epsilon_t$ is the random error with a zero mean and finite variance $\sigma^2$ at all times.  The single mean shift case has the structure
\[ 
\mu_t = 
\left\{ 
\begin{array}{l l}
	\Delta_1,     & \quad 1 \leq  t \leq \tau,  \\
	\Delta_2,     & \quad \tau < t  \leq N      \\
\end{array} \right. .
\]
for an unknown changepoint time $\tau \in \{2, \ldots, N \}$ (time 1 is not allowed to be a changepoint). The quantity $\Delta_2 - \Delta_1$ is the size of the mean shift.  In the multiple changepoint case where possibly more than one changepoint exists, we let $m$ denote the unknown number of changepoints and write
\[ 
\mu_t = 
\left\{ 
\begin{array}{l l}
	\Delta_1,     & \quad 1   \leq  t  \leq \tau_1, \\
	\Delta_2,     & \quad \tau_1  < t \leq \tau_2, \\
	~\vdots      & ~~~~~~~~~~\vdots            \\
	\Delta_{m+1}, & \quad \tau_m  < t  \leq  N.    \\
\end{array} \right. .
\]
This structure allows each of the $(m+1)$ ``regimes'' of the series to have a distinct mean.

\section{The Deleterious Effects of Changepoints}

\begin{figure}[t]
	\caption{The Tuscaloosa temperature series in degrees Celsius.  A fitted simple linear regression (blue dotted line) has a slightly negative trend slope.  This slope becomes significantly positive when two mean shifts, occurring in 1939 and 1957, are incorporated into the fit.}
	\centering
	\noindent\includegraphics[width=1\textwidth]{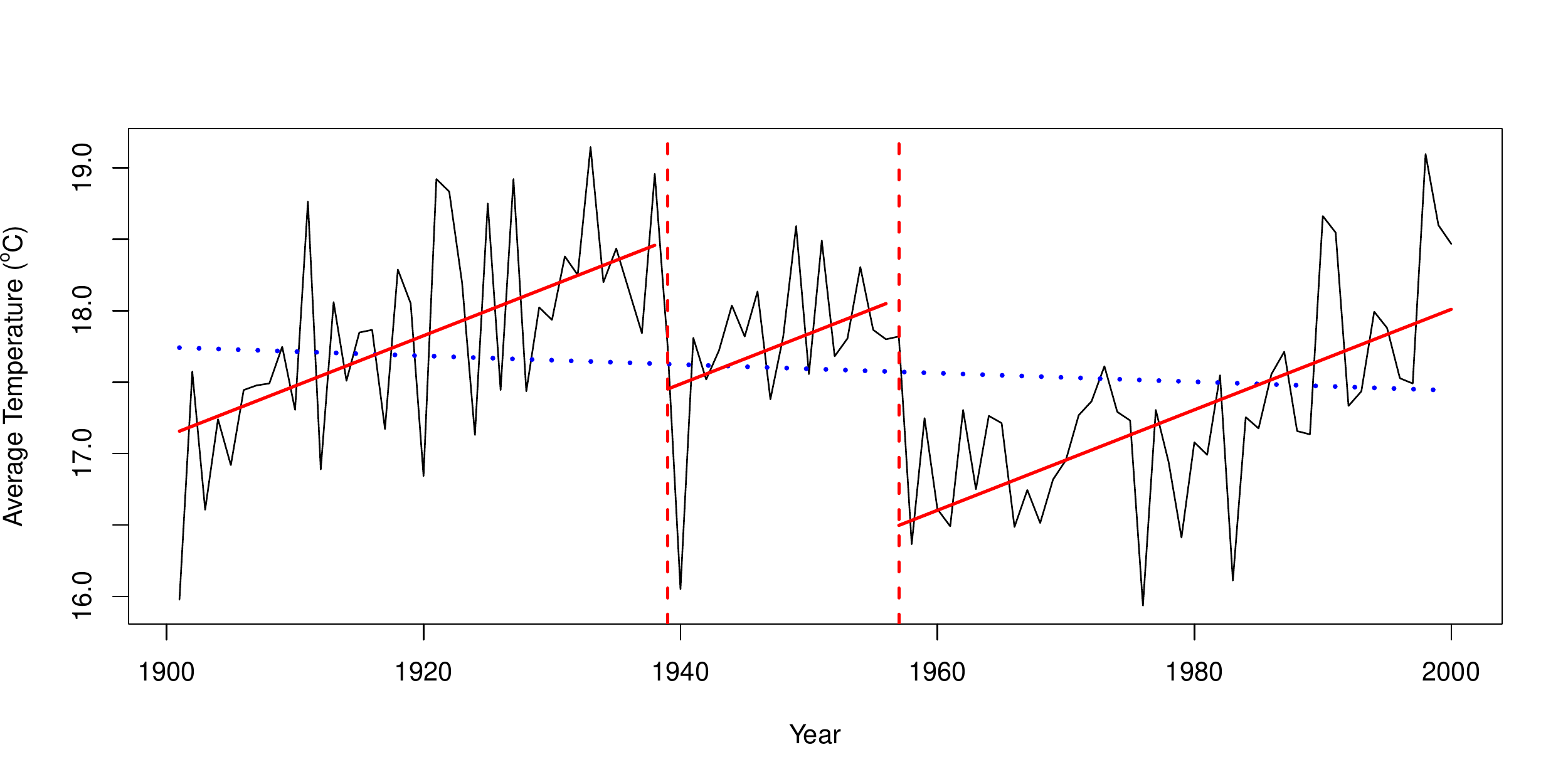}
	\label{fig:tuscaloosa}
\end{figure}

We now illustrate the havoc that changepoints wreak on trend estimation with an analysis of a century of annual temperatures from Tuscaloosa, Alabama, observed from 1901-2000.  This series is plotted in Figure \ref{fig:tuscaloosa} along with two regression fits that are explained below. Our analysis examines the linear change rate of the series to assess global warming.

Suppose that we were naive and neglected all changepoint issues.  A ground zero approach to the problem fits the simple linear regression $\mu_t = \alpha + \beta t$ to the series. Here, $\alpha$ is a location parameter and $\beta$ is the linear trend slope. Fitting this model provides the estimates 
\[
\hat{\alpha} =   17.74^\circ C, \quad
\hat{\beta}  =   -0.302^\circ C/{\rm Century}.
\]
We will not concern ourselves with standard errors for these estimates here; rather, our point is that the trend slope estimate $\hat{\beta}$ is negative, suggesting that Tuscaloosa has cooled (albeit slightly) during the century.  This fit is depicted in the dotted line in Figure \ref{fig:tuscaloosa}.  Cooling is not necessarily absurd here in a global warming era: the Earth has not warmed uniformly by spatial location and the Southeastern United States is known to have enjoyed somewhat of a `warming hole' during this era.   

In contrast, suppose that we fit a model with both a linear trend and an unknown number of mean shifts --- say $\mu_t = E[X_t] = \Delta_i + \beta t$ for times $t$ in the $i$th regime (the $m$ changepoints partition the time indices $1, 2, \ldots, N$ into $m+1$ distinct regimes).  The trend slope is constant across all regimes. The model is fitted by a penalized Gaussian likelihood with first order autoregressive errors, explained further in Section 5.  This fit estimates the number of changepoints $m$ and their locations $\tau_1, \ldots, \tau_m$ and is plotted in Figure 1.   The fit estimates two changepoints that occur in 1939 and 1957. While the fit improves on the simple linear fit, something sinister has happened:  the trend slope sign has reversed, becoming a whopping $3.52^\circ C$ per century.  This entails significant warming!  With the Tuscaloosa series, the changepoint times also make sense:  a station relocation is listed in the extended station logs in 1939 and two gauge changes during 1956 (station logs, called metadata, are often unavailable). 

The Tuscaloosa series is not pathological; indeed, a typical century-long temperature series has multiple changepoints during its record, with some shifts being a degree or two in magnitude.  These shifts often move the series in opposite directions and make accurately estimating a long-term trend, which is often less than the magnitude of one of the mean shifts over the record period, a daunting task. Due to the number of recording stations available, the situation is not entirely hopeless; however, climatologists now realize that trends computed for a single station are untrustable when changepoint effects are neglected.  In many changepoint problems, the fundamental issue lies with estimating the number of changepoints and where they occur; once this is done, most statistical procedures proceed in a straightforward manner.

\section{Changepoint Uses}

\subsection{Homogeneity Adjustments}

Perhaps the chief use of changepoint methods in climatology is to homogenize series. Climate homogenizers seek to adjust series for any man-made effects by subtracting/adding estimates of the $\Delta_i$'s to the series. This rids the series of the man-made effects induced by gauge changes, station relocations, or other measurement change methods, leaving a record that can be attributed only to natural forcings.  Climatologists prefer to leave changepoint effects from natural causes in the series, deeming it part of the `true record'.   Adjusting a series for an excessive number of changepoints risks removing too much variation in the record, making other fluctuations seem more significant than they truly are. 

To homogenize series, climatologists often compare a series for the location under study, called the target series, to a series of the same quantity collected at a nearby recording station that experiences similar weather, called a reference series.  Often, many reference series are used to make conclusions, sometimes as many as 40 in practice.  If $\{ X_t \}$ denotes the target series and $\{ Y_t \}$ the reference series, climatologists often analyze the difference $\{ X_t - Y_t \}$ for changepoints.  If the reference series matches the target series well, it should have similar fluctuations (weather) as the target series. The subtraction should hence eliminate most naturally occurring fluctuations. However, if the target series has a shift induced my a man-made changepoint (say a station relocation or a gauge change), then it will also be a changepoint in the target minus reference series.  

Unfortunately, by making target minus reference comparisons, changepoints in either the target or reference series become changepoints in the target minus reference series.  This in essence doubles the number of changepoints in the series, complicating matters for the homogenizer. Modern homogeneity methods combat this issue by examining many target minus reference series:  if the target minus reference series flags the same changepoint time for ten distinct station references, then evidence suggests that the changepoint time is attributable to the target series.  There exist stations where multiple good reference series do not exist; the North Atlantic Basin tropical cyclone analysis in the next subsection has no references series whatever.

\subsection{Stationarity Assessments}

Another use of changepoint methods lies with stationarity assessments. A common null hypothesis for climatologists is that some aspect of climate is static (stationary). Changepoint methods can be used to assess this hypothesis: if a multiple changepoint procedure flags one or more changepoints, then the record is deemed non-stationary.  The form of the mean $\mu_t=E[X_t]$ is not important here; linear trends, sinusoidal components, and even hinge shaped structures have arisen in practice.  In general, a multiple mean shift changepoint procedure will flag changepoints that attempt to follow the mean of the series.  For example, if the series has an increasing linear trend, then a mean shift changepoint routine will attempt to approximate this by flagging multiple changepoints that result in an increasing staircase.

\begin{figure}[t]
	\caption{Tropical Cyclone Counts in the North Atlantic Ocean since 1970. The prominent changepoint in 1995 indicates that the record is non-stationary, with the region experiencing enhanced storm activity since this time.}
	\centering
	\noindent\includegraphics[width=1\textwidth]{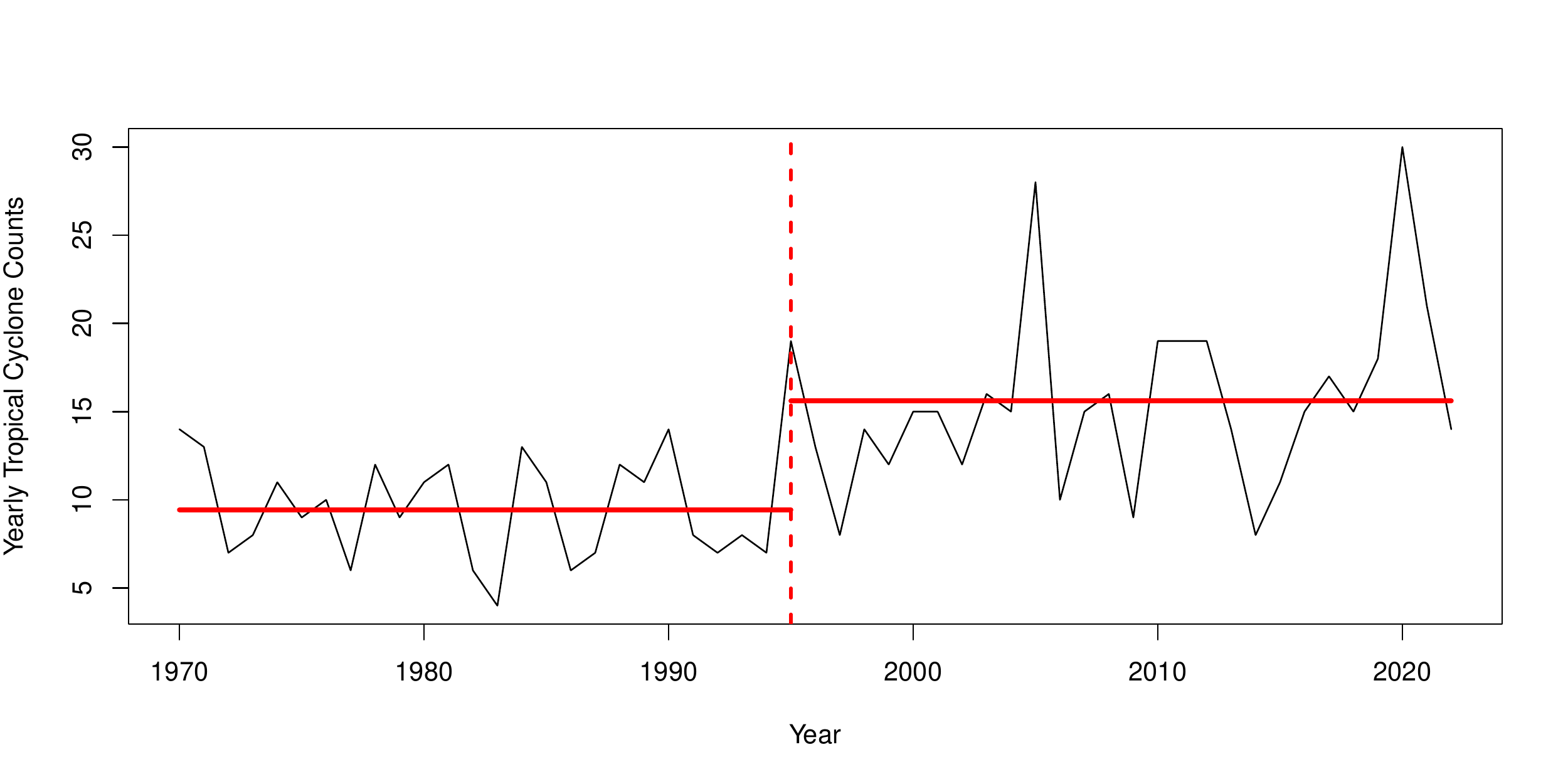}\\
	\label{fig:tropical}
\end{figure}

As an example of a stationarity check, we consider the North Atlantic Ocean Basin's tropical cyclone counts for the 53-year period 1970-2022. These are plotted in Figure \ref{fig:tropical} against a mean shift changepoint fit that is described below.

Annual tropical cyclone counts in the North Atlantic Basin are well modeled by a Poisson marginal distribution.  Hence, a reasonable model posits that the cyclone count for year $t$, denoted by $X_t$, has a Poisson distribution with parameter $\lambda_t$.  Since $\mu_t = \lambda_t$ in this case, we conduct a multiple changepoint analysis with
\[ 
\lambda_t = 
\left\{ 
\begin{array}{l l}
	r_1,     & \quad 1   \leq t  \leq \tau_1,  \\
	r_2,     & \quad \tau_1 < t  \leq \tau_2,  \\
	~\vdots & ~~~~~~~~~~\vdots                 \\
	r_{m+1}, & \quad \tau_m < t  \leq  N       \\
\end{array} \right. .
\]
The year-to-year storm counts are taken to be statistically independent; indeed, there is very little predictive power in a tropical cyclone count forecast even 12 months in advance (annual conferences are held each year in May, just prior to the start of hurricane season, where various modelers release their annual count forecast).

The multiple changepoint fit estimates one changepoint in 1995. The estimated mean of the series is plotted in Figure \ref{fig:tropical} against the storm counts.  Clearly, we have entered into an era of enhanced tropical cyclone activity circa the mid 1990s, a concerning aspect for Eastern Coastal residents.

\section{A Single Mean Shift}

The rest of this article explains a few basic mean shift changepoint techniques.  This section examines the simplest case where there is either one or no changepoints in the record --- the at most one changepoint (AMOC) setting.  Our analysis assumes that $\{ X_t \}$ is independent and that there are no trends in the series.  

If time $k$ were in truth a mean shift, the segment means $k^{-1}\sum_{t=1}^k X_t$ and $(N-k)^{-1}\sum_{t=k+1}^N X_t$ should be statistically different from zero. Scaling this difference further to converge to a proper limiting distribution leads to the cumulative sum (CUSUM) statistic at time $k$:
\[
\mbox{CUSUM}(k) = \frac{
	\sum_{t=1}^k X_t - \frac{k}{N}\sum_{t=1}^N X_t}
{\sigma \sqrt{N}},
\]
where $\sigma^2=\mbox{Var}(X_t)$.  Since we do not know $\sigma^2$, we replace it with the no-changepoint null hypothesis estimate
\[
\hat{\sigma}^2 = 
\frac{\sum_{t=1}^N (X_t -\overline{X})^2}{N-1},
\]
where $\overline{X} = N^{-1}\sum_{t=1}^N X_t$ is the sample mean of the entire series. Because the changepoint time is unknown, it is estimated as the location of the absolute largest CUSUM statistic:
\[
\hat{\tau}= \argmax_{2 \leq k \leq N} 
|{\rm CUSUM}(k)|.
\]

Asymptotic percentiles as $N \rightarrow \infty$ are related to Brownian bridge processes. Table \ref{tab:cusum_dist_tab} below lists some commonly used percentiles obtained by simulation. 
Many other AMOC tests exist in the literature; this said, the CUSUM test has been a changepoint staple for years.

\begin{table}
	\caption{Asymptotic Critical Values for CUSUM Statistics}
	\label{tab:cusum_dist_tab}
	\begin{center}
		\begin{tabular}{ c  c }
			\hline
			Percentile    &    Critical Value  \\
			\hline
			\hline
			90.0\%      & 1.224       \\
			95.0\%      & 1.358       \\
			97.5\%      & 1.480       \\
			99.0\%      & 1.628       \\
			\hline
		\end{tabular}
	\end{center}
\end{table}

\section{Multiple Changepoints}

As we have seen with the Tuscaloosa series, climate time series can have more than one changepoints.  Multiple changepoint techniques have been heavily researched in recent years, with various techniques being proposed. This section examines two of them:  binary segmentation and penalized likelihoods.

\subsection{Binary Segmentation and its Variants}

Binary segmentation methods essentially turn an AMOC procedure into a multiple changepoint technique.  The process is simple: one examines the entire series for the existence of a single changepoint.  If none is found, the procedure is stopped and no chagepoints are declared and the entire series is declared changepoint free.  Should a changepoint be flagged, the series is split about the estimated changepoint time and the two series subsegments (before and after the changepoint time) are scrutinized for additional single changepoints.  The procedure is repeated iteratively, stopping when all series subsegments are declared changepoint free. Binary segmentation methods can euse any AMOC test --- it need not be a CUSUM procedure.

Binary segmentation calculations proceed very rapidly in comparison to some other multiple changepoint techniques.  In fact, the calculations are almost instantaneous, even for long series.  Unfortunately, binary segmentation is one of the worst performing multiple changepoint procedures, often being fooled when multiple mean shifts move the series in different directions. Simulations reinforce this point.

Attempted fixes to binary segmentation include wild binary segmentation (WBS) and its related procedures.  WBS examines randomly selected subintervals for a single changepoint with a CUSUM AMOC test and then combines the conclusions from all random subintervals to estimate the multiple changepoint configuration.  With its current settings, WBS is an aggressive procedure that often overestimates the number of changepoints, especially when few changepoints in truth exist.  Research on these methods continues today.

\subsection{Penalized Likelihoods}

Another way to estimate a multiple changepoint configuration uses likelihoods that are penalized to prevent fitting an excessive number of changepoints. Changepoint configurations were estimated in this manner earlier for the Tuscaloosa temperature data and the North Atlantic tropical cyclone counts.

Suppose that $m$ changepoints occur at the times $\tau_1, \ldots , \tau_m$.  The statistical likelihood of such a model will be denoted by $L(m; \tau_1, \ldots, \tau_m)$.  For the Tuscaloosa data, this likelihood was based on a Gaussian time series model with first order autoregressive errors; for the North Atlantic tropical cyclone counts, the likelihood is based on independent Poisson counts.

As more changepoints are added to the model, the resulting likelihood becomes bigger, or  equivalently, $-2\ln (L(m; \tau_1, \ldots, \tau_m))$ becomes smaller.  However, once the optimal model is reached, adding additional changepoints to the fit does little to improve its likelihood.  A penalized likelihood procedure adds a penalty to the negative log likelihood for having $m$ changepoints at the times $\tau_1, \ldots , \tau_m$.  We denote the penalty by $P(m; \tau_1, \ldots, \tau_m)$.  The penalized likelihood estimated changepoint configuration is the model that minimizes
\[
-2\ln(L(m;\tau_1, \ldots, \tau_m)) + 
P(m; \tau_1, \ldots, \tau_m).
\]

Many different penalty types have been proposed. One that seems to work well in mean shift problems is the BIC penalty $P(m; \tau_1, \ldots , \tau_m)=m\ln(N)$.  The classical AIC penalty $P(m; \tau_1, \ldots, \tau_m)=2m$ tends to estimate too many changepoints. There exist more exotic penalties that depend on the times of the changepoints (one such is a minimum description length).

While penalized likelihood methods perform better than binary segmentation in simulations, they are much more computationally intensive.  In particular, the objective function above is highly non-convex, making locating its minimum unwieldy.  An exhaustive search over all changepoint configurations requires calculations to fit $2^{N-1}$ distinct models (this tally is not $2^N$ since time 1 cannot be a changepoint).
This exceeds one billion model fits when $N \geq 32$.  To combat this issue, researchers have devised genetic algorithms to optimize the penalized likelihood.  Genetic algorithms are intelligent random walk searches that are unlikely to fit models at suboptimal changepoint configurations.   These techniques rely on random generations of changepoint configurations that evolve according to elements of genetic fitness and mutation.  While many "off the shelf" genetic algorithms can now reliably handle the problem, they often take hours of computing time.

\section{Comments and Conclusions}

Changepoint methods have multiple uses in climatology, including stationary checks and record homogenization.  There are still many open problems in the area, especially in the multiple changepoint setting, and statisticians are needed to help develop the methods and analyze the data.

For additional reading on penalized likelihood multiple changepoint methods, see \cite{Davis_etal_2006}, \cite{Shi_etal_2021}, and \cite{JCL-CET}.  The reference \cite{Fry-2014} develops wild binary segmentation methods; genetic algorithm development for multiple changepoint problems is addressed in \cite{Lu_Lund_Lee}.  A more detailed overview on changepoint methods in climatology is contained in \cite{JCL_Review}.  A deeper analysis of the tropical cyclone data is contained in \cite{Robbins_etal_JASA}; the Tuscaloosa data is analyzed further in \cite{Lu_Lund_Canadian}.

\bibliographystyle{unsrt}

\end{document}